\documentclass[aps,prl,twocolumn,superscriptaddress,sort&compress]{revtex4-1}%
\usepackage[T1]{fontenc}
\usepackage{lmodern}
\usepackage{eurosym}
\usepackage{amsfonts}
\usepackage{amssymb}
\usepackage{amsmath}
\usepackage{floatrow}
\usepackage[position=t,singlelinecheck=off,caption=false]{subfig}
\usepackage{graphicx}
\usepackage{xcolor}%
\usepackage{booktabs}
\usepackage{adjustbox}
\usepackage[hyperfigures=true,bookmarksnumbered=true,bookmarksopen=true,bookmarks=true,linkcolor=blue,urlcolor=blue,citecolor=blue,colorlinks=true,pdfhighlight=/O,pdfstartview={XYZ null null 1}]{hyperref}
\graphicspath{{}{images/}}
\providecommand{\U}[1]{\protect\rule{.1in}{.1in}}
\providecommand{\U}[1]{\protect\rule{.1in}{.1in}}
\def\HG#1 {\emph{\color{blue}#1}}

\begin{document}
\title{Anomalous Hall effect in non-collinear antiferromagnetic antiperovskite Mn$_{3}$Ni$_{1-x}$Cu$_{x}$N}
\author{K. Zhao}
\affiliation{Experimentalphysik VI, Center for Electronic Correlations and Magnetism, University of Augsburg, 86159 Augsburg, Germany}
\author{T. Hajiri}
\affiliation{Department of Materials Physics, Nagoya University, Nagoya 464-8603, Japan}
\author{H. Chen}
\affiliation{Department of Physics, Colorado State University, Fort Collins, CO 80523-1875, USA}
\author{R. Miki}
\affiliation{Department of Materials Physics, Nagoya University, Nagoya 464-8603, Japan}
\author{H. Asano}
\affiliation{Department of Materials Physics, Nagoya University, Nagoya 464-8603, Japan}
\author{P. Gegenwart}
\affiliation{Experimentalphysik VI, Center for Electronic Correlations and Magnetism, University of Augsburg, 86159 Augsburg, Germany}

\begin{abstract}
We report the anomalous Hall effect (AHE) in antiperovskite Mn$_{3}$NiN with substantial doping of Cu on the Ni site (i.e. Mn$_{3}$Ni$_{1-x}$Cu$_{x}$N), which stabilizes a noncollinear antiferromagnetic (AFM) order compatible with the AHE. Observed on both sintered polycrystalline pieces and single crystalline films, the AHE does not scale with the net magnetization, contrary to the conventional ferromagnetic case. The existence of the AHE is explained through symmetry analysis based on the $\Gamma_{\rm 4g}$ AFM order in Cu doped Mn$_{3}$NiN. DFT calculations of the intrinsic contribution to the AHE reveal the non-vanishing Berry curvature in momentum space due to the noncollinear magnetic order. Combined with other attractive properties, antiperovskite Mn$_{3}$AN system offers great potential in AFM spintronics.
\end{abstract}
\maketitle

Empirically, the anomalous Hall effect (AHE) in conventional ferromagnetic metals is proportional to the net magnetization ~\cite{AHE2010RMP} and was expected to vanish in antiferromagnets. However, it is now well understood that the existence of the AHE is constrained by symmetry and is not necessarily incompatible with AFM order. This principle has been exemplified by the large AHE predicted and realized in various noncollinear AFMs, such as hexagonal Mn$_{3}$Sn/Ge and cubic Mn$_{3}$Ir/Pt, etc. where the Mn atoms form a kagome lattice along different crystalline planes~\cite{Mn3Ir2014PRL, Mn3Sn2014EPL, Mn3Sn2015Nature, Mn3Ge2016PRA, Mn3Ge2016SA, Mn3Sn2017NJP, Mn3Sn2017NM, Mn3Pt2018NE}. In these materials, geometric frustration and local symmetries of the magnetic atoms together lead to a chiral noncollinear magnetic order with the essential symmetry breaking for finite AHE, which is also reflected by the small net moment in certain crystalline directions.
%Although the net size of the AHE in real materials depends on both intrinsic and extrinsic contributions, the latter of which are difficult to determine quantitatively,
The large AHE in the hexagonal Mn$_{3}$Sn/Ge was argued to be due to Weyl points near the Fermi energy, which can be viewed as monopoles of Berry curvature in momentum space~\cite{Mn3Ir2014PRL, Mn3Sn2014EPL, Mn3Sn2015Nature, Mn3Ge2016PRA, Mn3Ge2016SA, Mn3Sn2017NJP, Mn3Sn2017NM, Mn3Pt2018NE}. Technologically, the robust AHE at room temperature in these systems makes them attractive for potential applications in spintronics solely based on AFM~\cite{Review2018NP}. Meanwhile, it is interesting to search for similar anomalous Hall antiferromagnets (AHE AFMs) beyond binary Mn compounds, which may have other superior properties.

\begin{figure}[t]
 \includegraphics[width=1.0\columnwidth]{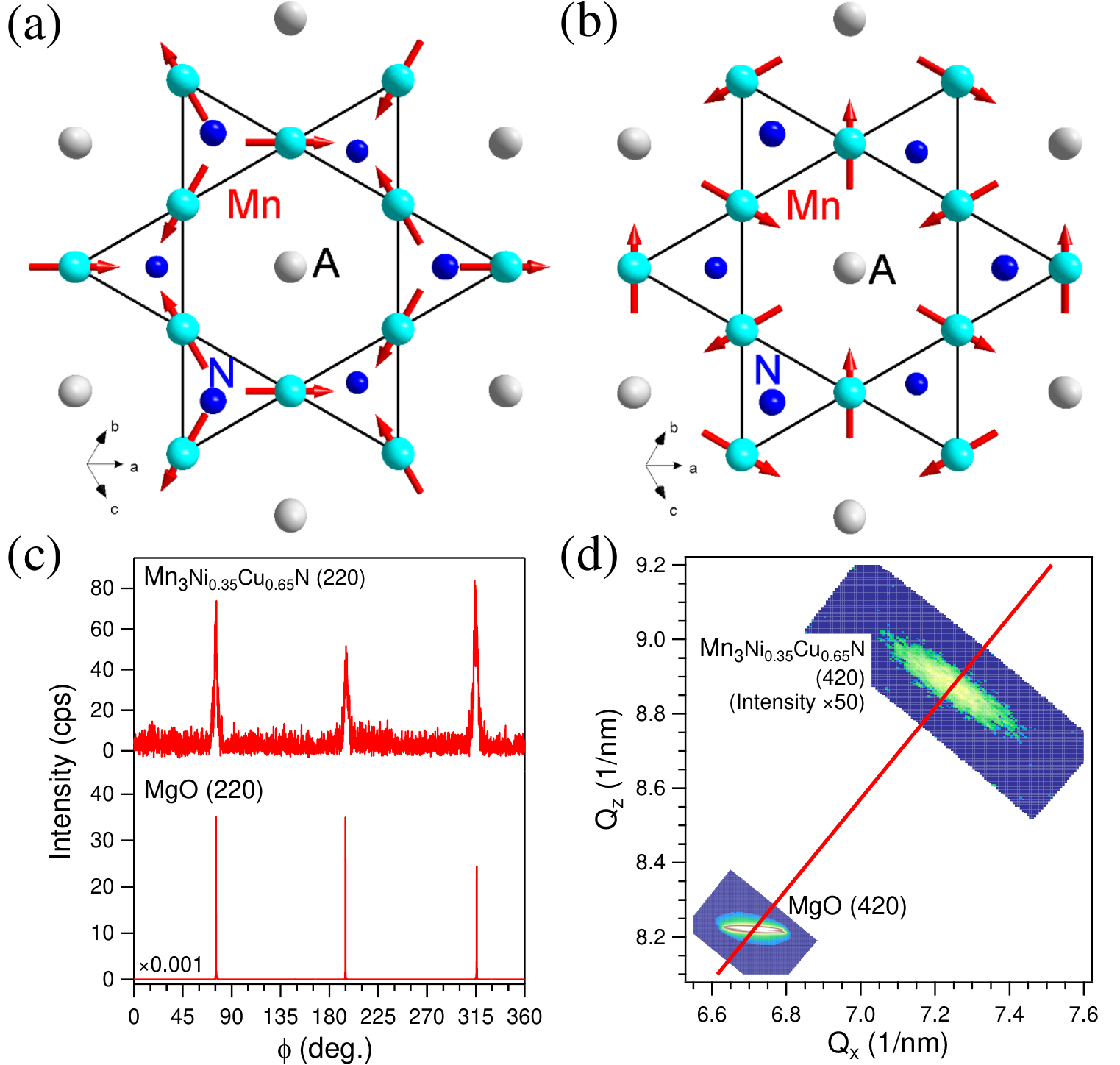}
 \caption{(Color online) \textbf{Magnetic and crystal structures of Mn$_{3}$Ni$_{1-x}$Cu$_{x}$N.} (a)-(b) The Mn kagome lattice with noncollinear AFM structures termed $\Gamma_{\rm 4g}$ and $\Gamma_{\rm 5g}$ in Mn$_{3}$AN, respectively. (c) In-plane X-ray diffraction data of Mn$_{3}$Ni$_{0.35}$Cu$_{0.65}$N 100 nm film grown on MgO (111) substrate, with three fold symmetry of (220) peak, (d) Reciprocal space map around (420) reflection, with the solid line as the relaxation line.}
 \label{fig:1}
 \end{figure}

\begin{figure*}[t]
 \includegraphics[width=1.03\columnwidth]{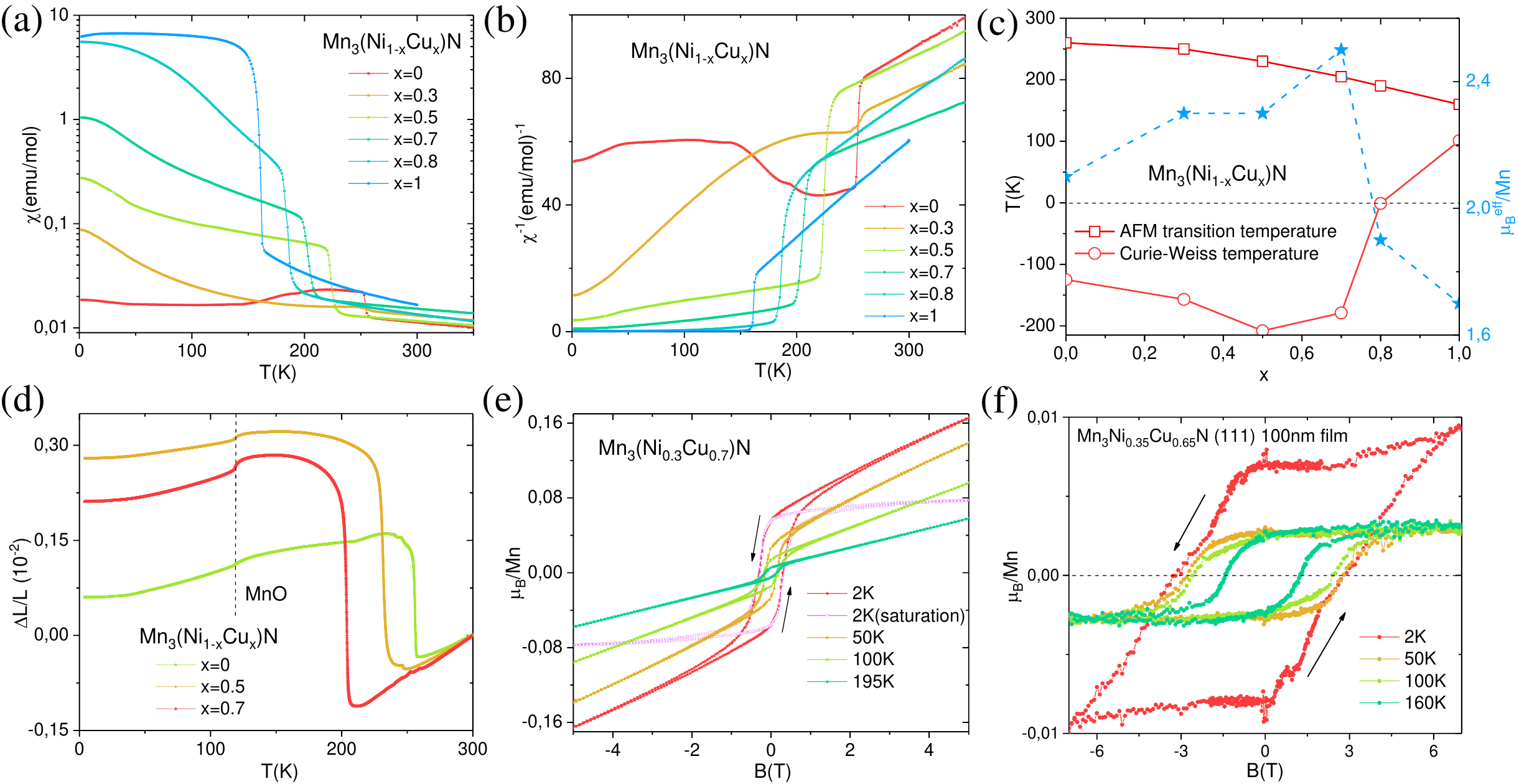}
 \caption{(Color online) \textbf{Magnetic properties and thermal expansion of Mn$_{3}$Ni$_{1-x}$Cu$_{x}$N.} (a)-(b) DC magnetic susceptibility and inverse susceptibility measured in H=500 Oe in Mn$_{3}$Ni$_{1-x}$Cu$_{x}$N with several different Cu doping levels $x$ from 2K to 350K under field cooling. (c) Phase diagram of Mn$_{3}$Ni$_{1-x}$Cu$_{x}$N ($0<x<1$), including the experimental AFM transition temperature, the Curie-Weiss temperature, and the effective moment vs. Cu doping level. The latter two are obtained from Curie-Weiss fitting of the high-temperature susceptibilities. (d) Temperature-dependent relative length change [$\Delta$L/L=[L(T)-L(300K)]/L(300K)] for Mn$_{3}$Ni$_{1-x}$Cu$_{x}$N for $x=$0, 0.5, and 0.7, obtained upon increasing temperature. (e) Isothermal magnetization $M(H)$ for Mn$_{3}$Ni$_{0.3}$Cu$_{0.7}$N at various temperatures, with saturation moment at 2K also shown(see text). (f) Out of plane $M(H)$ for Mn$_{3}$Ni$_{0.35}$Cu$_{0.65}$N 100nm film on MgO (111) substrate at various temperatures.}
 \label{fig:2}
\end{figure*}

Anti-perovskite manganese nitrides Mn$_{3}$AN, have received a lot of attention due to a number of unusual physical properties, such as giant negative thermal expansion~\cite{Mn3GaN2005APL, Mn3AN2014NTE}, temperature independent resistivity ~\cite{Mn3NiCuN2011APL}, and barocaloric effects~\cite{Mn3GaN2015NM, Mn3NiN2018PRX}. All of these phenomena are closely related to the first order phase transition from a high temperature paramagnetic cubic phase into a low temperature noncollinear AFM phase with the so-called $\Gamma_{\rm 4g}$ and $\Gamma_{\rm 5g}$ magnetic structures. It is interesting to notice that the Mn atoms form a 1/4-depleted fcc lattice in these manganese antiperovskites, similar to that in the AHE AFM Mn$_3$Ir and Mn$_3$Pt, whose (111) planes are kagome lattices as shown in Fig. 1(a) and (b)~\cite{Mn3AN1978JPSJ, Mn3NiN2013JAP}. Therefore one would expect there to be AHE if the AFM order of the Mn moments is similar to these binary compounds.

In this Rapid Communication, we demonstrate that the AHE could indeed be realized in manganese nitrides, by studying the Mn$_{3}$NiN system~\cite{Mn3NiCuN2011APL, Mn3AN2014NTE, Mn3NiN2013JAP, Mn3NiCuN2015JAP}. We first found that in stoichiometric  Mn$_{3}$NiN, a weak signal of AHE is observed below the noncollinear AFM ordering temperature of 180K, yet disappeared as the temperature further decreases. Using magnetometry and thermal expansion measurements, we then show that through substantial doping of Cu substituting Ni, the noncollinear AFM order is still preserved in Mn$_{3}$Ni$_{1-x}$Cu$_{x}$N polycrystals. Most interestingly, the AHE is enhanced and becomes more robust against temperature after Cu doping. The AHE also does not scale with the net magnetization, contrary to the ferrimagnetic Mn$_{3}$CuN system~\cite{Mn3CuN2001SSC, Mn3CuN2017PRB}. We have also successfully grown single-crystalline Mn$_{3}$Ni$_{1-x}$Cu$_{x}$N films, which exhibit large AHE as well. Finally, we analyze the symmetry origin of the AHE in the $\Gamma_{\rm 4g}$ state of Mn$_{3}$NiN, complemented by first-principles density functional theory (DFT) calculations revealing the momentum space Berry curvature contributing to the AHE.

\begin{figure*}[t]
 \includegraphics[width=1.02\columnwidth]{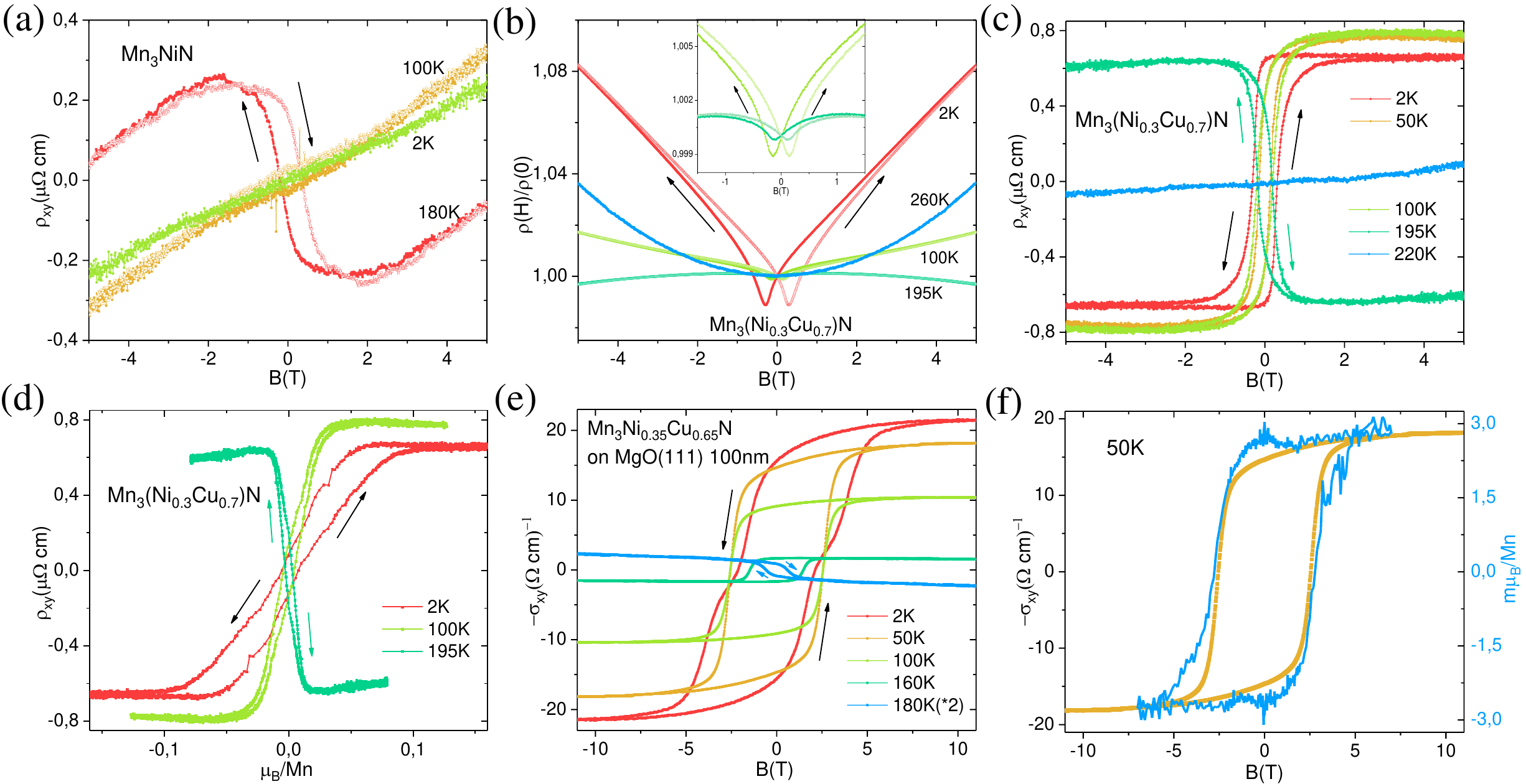}
 \caption{(Color online) \textbf{Transport properties including the AHE in Mn$_{3}$Ni$_{1-x}$Cu$_{x}$N.} (a) Hall resistivity results from a sintered specimen of Mn$_{3}$NiN at 180K, 100K, and 2K, respectively. (b) Field dependence of the longitudinal resistivity $\rho(H)/\rho(0)$ of Mn$_{3}$Ni$_{0.3}$Cu$_{0.7}$N at various temperatures, with the low field region at 195K and 100K enlarged in the inset. (c) Field dependence of the Hall resistivity $\rho_{xy}$ of Mn$_{3}$Ni$_{0.3}$Cu$_{0.7}$N at various temperatures, with the AHE clearly observed below 205K. (d) Dependence of $\rho_{xy}$ on magnetization of Mn$_{3}$Ni$_{0.3}$Cu$_{0.7}$N at 195K, 100K, and 2K, respectively. (e) The anomalous Hall conductivity (AHC) versus field measured in single crystalline Mn$_{3}$Ni$_{0.35}$Cu$_{0.65}$N film on MgO (111) substrate at various temperatures (see text). (f) Comparison of the Hall conductivity and isothermal magnetization of (111) 100nm film at 50K.}
 \label{fig:3}
 \end{figure*}

The successful homogeneous solution of Mn$_{3}$Ni$_{1-x}$Cu$_{x}$N powders have been verified by X-ray diffraction in Fig. S1~\cite{Mn311SM}, consistent with previous results~\cite{Mn3NiCuN2011APL, Mn3AN2014NTE, Mn3NiCuN2015JAP}. Fig.~\ref{fig:1} (c) shows in-plane X-ray diffraction pattern of Mn$_{3}$Ni$_{0.35}$Cu$_{0.65}$N film. The out-of-plane X-ray diffraction pattern is shown in Fig. S2~\cite{Mn311SM, Mn3GaNfilm2013}. On the MgO (111) substrate, (111)-orientated films are epitaxially grown with three-fold in plane symmetry. Fig.~\ref{fig:1} (d) displays the reciprocal space map of Mn$_{3}$Ni$_{0.35}$Cu$_{0.65}$N film on MgO (111) substrate. The (420) peak is observed on the relaxation line, indicating no stain condition of our high quality film.

Figure~\ref{fig:2} (a) shows the temperature dependence of magnetic susceptibilities in field-cooling (FC) procedures under 500 Oe for Mn$_{3}$Ni$_{1-x}$Cu$_{x}$N specimens with $x=$0, 0.3, 0.5, 0.7, 0.8, and 1, respectively. The pure Mn$_{3}$NiN shows the AFM transition at 260K, consistent with previous results~\cite{Mn3NiCuN2011APL, Mn3AN2014NTE, Mn3NiCuN2015JAP}. 
%The Cu doping continuously decreases the temperature of the AFM transition, with transition temperature $T_N=$ 260 K, 250 K, 230 K, 205K, 190 K, and 160 K, respectively, and increases the low temperature $\chi(T)$, until the system becomes ferrimagnetic for $x>$0.7, which is similar to pure Mn$_{3}$CuN. 
The Cu doping continuously decreases the temperature of the AFM transition and increases the low temperature $\chi(T)$, until the system becomes ferrimagnetic which is similar to pure Mn$_{3}$CuN, with transition temperature $T_N=$ 260 K, 250 K, 230 K, 205K, 190 K, and 160 K, respectively. Above $T_{N}$, the samples are paramagnetic, and the inverse susceptibility $\chi^{-1}(T)$ can be fitted to the Curie-Weiss formula [Fig. 2(b)]. Fig. 2(c) summarizes the phase diagram of Mn$_{3}$Ni$_{1-x}$Cu$_{x}$N system, by plotting the $T_N$ versus Cu doping level. Also shown in the plot are the Curie-Weiss temperature and the effective moment obtained from the Curie-Weiss fitting. Similar to the $x=0$ case, the $x=$0.3, 0.5, and 0.7 samples exhibit negative Curie-Weiss temperature, suggesting AFM-type interaction, which becomes nearly zero for $x=$0.8, and about +100K for the ferrimagnetic Mn$_{3}$CuN \cite{Mn3CuN2001SSC}. The effective moment $\mu_{eff}$ sharply decreases when $x>$0.7, consistent with the change of the ground state.

As already verified by neutron diffraction experiments in various Mn$_{3}$AN systems~\cite{Mn3CuGeN2008PRB, Mn3CuGeN2010PRB, Mn3ZnN2012PRB, Mn3NiN2013JAP}, the negative thermal expansion there is due to the volume expansion from the paramagnetic cubic phase to the noncollinear AFM phase, while there is no such effect from the paramagnetic-ferrimagnetic transition in Mn$_{3}$CuN. As shown in Fig.~\ref{fig:2} (d), for Mn$_{3}$Ni$_{1-x}$Cu$_{x}$N with $x=$0, 0.5, and 0.7, all three specimens exhibit large volume expansion through a sharp transition at $T_N$. Moreover, the Cu doped samples show even larger volume expansion than pure Mn$_{3}$NiN. The weak anomaly at 120K in the same plot is due to the AFM transition of MnO impurities~\cite{MnO1958PR}, supported by the X-ray diffraction data in Fig.~S1. The anomaly around 250K comes from the capacitive dilatometer based measurement system~\cite{TE2012}. Together with the magnetic results, our thermal expansion measurements provide clear evidence on the persistence of the noncollinear AFM order in Mn$_{3}$Ni$_{1-x}$Cu$_{x}$N for $x<$0.7.

As predicted and observed in Mn$_{3}$Sn and Mn$_{3}$Pt~\cite{Mn3Ir2014PRL, Mn3Sn2014EPL, Mn3Sn2015Nature, Mn3Sn2017NJP, Mn3Pt2018NE}, a small remnant net magnetization is always accompanying the essential symmetry breaking that leads to AHE. Pure Mn$_{3}$NiN exhibits a small hysteresis in the isothermal magnetization $M(H)$ curve at 180K as shown in Fig.~S3(a) \cite{Mn311SM}. However, the hysteresis is suppressed as temperature decreases, consistent with a spin glass type behavior observed at 2K. As shown in Fig. S3 (b) and (c) with $x=$0.3 and 0.5, the coercive field is still large at 50K, similar to pure Mn$_{3}$NiN system at 100K~\cite{Mn311SM}. Doping Cu in Mn$_{3}$NiN can gradually stabilize the small ferromagnetic (FM) component and presumably the magnetic structure that hosts AHE. This is best illustrated for the Mn$_{3}$Ni$_{0.3}$Cu$_{0.7}$N specimen in Fig. 2(e). A small FM component appears below 205K, and keeps increasing as temperature decreases. After subtracting the $H$-linear component in the high field region, presumably due to MnO impurities, the saturation moment is about 0.006$\mu_{B}$/Mn at 195K, 0.05$\mu_{B}$/Mn at 50K, and 0.08$\mu_{B}$/Mn at 2K, much smaller than its effective moment, 2.5$\mu_{B}$/Mn from the Curie-Weiss fitting. The coercive field also increases with decreasing temperature and reaches 0.3T at 2K.

Figure S5 shows the out of plane isothermal magnetization $M(H)$ curves of Mn$_{3}$Ni$_{0.35}$Cu$_{0.65}$N (111) 100nm film after subtracting the large signal from MgO (111) substrate~\cite{Mn311SM}. A rather weak remnant magnetization is clearly observed in Fig. 2(f), with the saturation moment $\sim 0.003\mu_{B}$/Mn above 50K and 0.01$\mu_{B}$/Mn at 2K, due to symmetry-allowed spin canting in the noncollinear state. This weak moment has a similar size to that in Mn$_{3}$Sn and Mn$_{3}$Pt~\cite{Mn3Sn2015Nature, Mn3Pt2018NE}, and checks with our DFT calculations below. Its direction is expected to be perpendicular to the kagome plane if the magnetic order is indeed $\Gamma_{\rm 4g}$. Presumably due to the crystalline order of the (111) films enhancing the magnetocrystalline anisotropy leading to a (111)-directed weak moment, we find the (111) orientated film shows much larger coercive field-- 1.5T at 160K, 2.5T at 100K and 50K, than the polycrystalline Mn$_{3}$Ni$_{0.3}$Cu$_{0.7}$N in Fig. 2(e).
%The coercive field is also much larger by several hundred gauss that in Mn$_{3}$Sn single crystal and Mn$_{3}$Pt single crystalline film~\cite{Mn3Sn2015Nature, Mn3Pt2018NE}.

Next we focus on the transport properties of Mn$_{3}$Ni$_{1-x}$Cu$_{x}$N samples, in particular the AHE. Fig.~\ref{fig:3} (a) shows weak signal of the AHE at 180K in pure Mn$_{3}$NiN, consistent with the small FM component shown in Fig. S3. This signal rapidly decreases with temperature, leaving only normal Hall signal at 2K, with the carrier density estimated to be $1.4\times 10^{22}$/cm$^3$. Mn$_{3}$NiN (111) film exhibits similar behavior as shown in Fig. S5~\cite{Mn311SM}.

Above 205K, the magnetoresistance (MR) of Mn$_{3}$Ni$_{0.3}$Cu$_{0.7}$N specimen increases as $H^2$ as shown in Fig. 3(b), consistent with normal metal behavior. Below 205K, clear hysteresis appears with the same coercive field value as in the $M(H)$ curve. The clear positive MR at 2K and 100K is also similar to that in Mn$_{3}$Sn~\cite{Mn3Sn2017NM}. The slightly negative MR at 195K is likely due to strong magnetic fluctuations close to $T_N$.

Most interestingly, Fig. 3(c) displays large AHE in the AFM state of Mn$_{3}$Ni$_{0.3}$Cu$_{0.7}$N specimen, with the same coercive field as in the $M(H)$ and MR curves. There is also sign change of the ordinary Hall resistance between 195K and 100K, suggesting a transition from hole-type carriers above 195K to electron-type carriers below 100K. To further determine the origin of such large AHE, we plot the Hall resistance $\rho_{xy}$ versus the magnetization $M$ in Fig.~\ref{fig:3} (d). Similar to the arguments used for decomposing different contributions to the AHE in Mn$_{3}$Sn~\cite{Mn3Sn2015Nature} and Mn$_{3}$Ge~\cite{Mn3Ge2016PRA}, the hysteresis with a sharp sign change in $\rho_{xy}(M)$ indicates the AFM-origin of the AHE. In contrast, $\rho_{xy}$ of the sintered Mn$_{3}$CuN in Fig. S6~\cite{Mn311SM} is dominated by a contribution proportional to $M$.

Further, the high quality epitaxial Mn$_{3}$Ni$_{0.35}$Cu$_{0.65}$N film on MgO (111) substrate enables us to extract the anomalous Hall conductivity (AHC) from the resistivity data, which can be more direcly compared with theory. According to the longitudinal and Hall resistance data in Fig. S7~\cite{Mn311SM}, the Hall conductivity $\sigma_{xy}$ is estimated using $\sigma_{xy} = - \rho_{xy} / \rho_{xx}^{2}$, which exceeds 20 ($\Omega$ cm)$^{-1}$ at 2K in Fig.~\ref{fig:3} (e). Similar to the powder case, the switching of AHC is very sharp, except for an additional anomaly at 2K. Both magnetization and the Hall conductivity of the (111) orientated film exhibit similar coercive field, as shown in Fig. 3(f) at 50K, indicating the large AHC is directly related with the noncollinear AFM order. The (111) orientated film also shows the sign change of the ordinary Hall signal in Fig. 3(e), which has not been observed in Mn$_{3}$Sn and Mn$_{3}$Pt~\cite{Mn3Sn2015Nature, Mn3Pt2018NE}, and deserves future study.

\begin{figure}[t]
 \includegraphics[width=1\columnwidth]{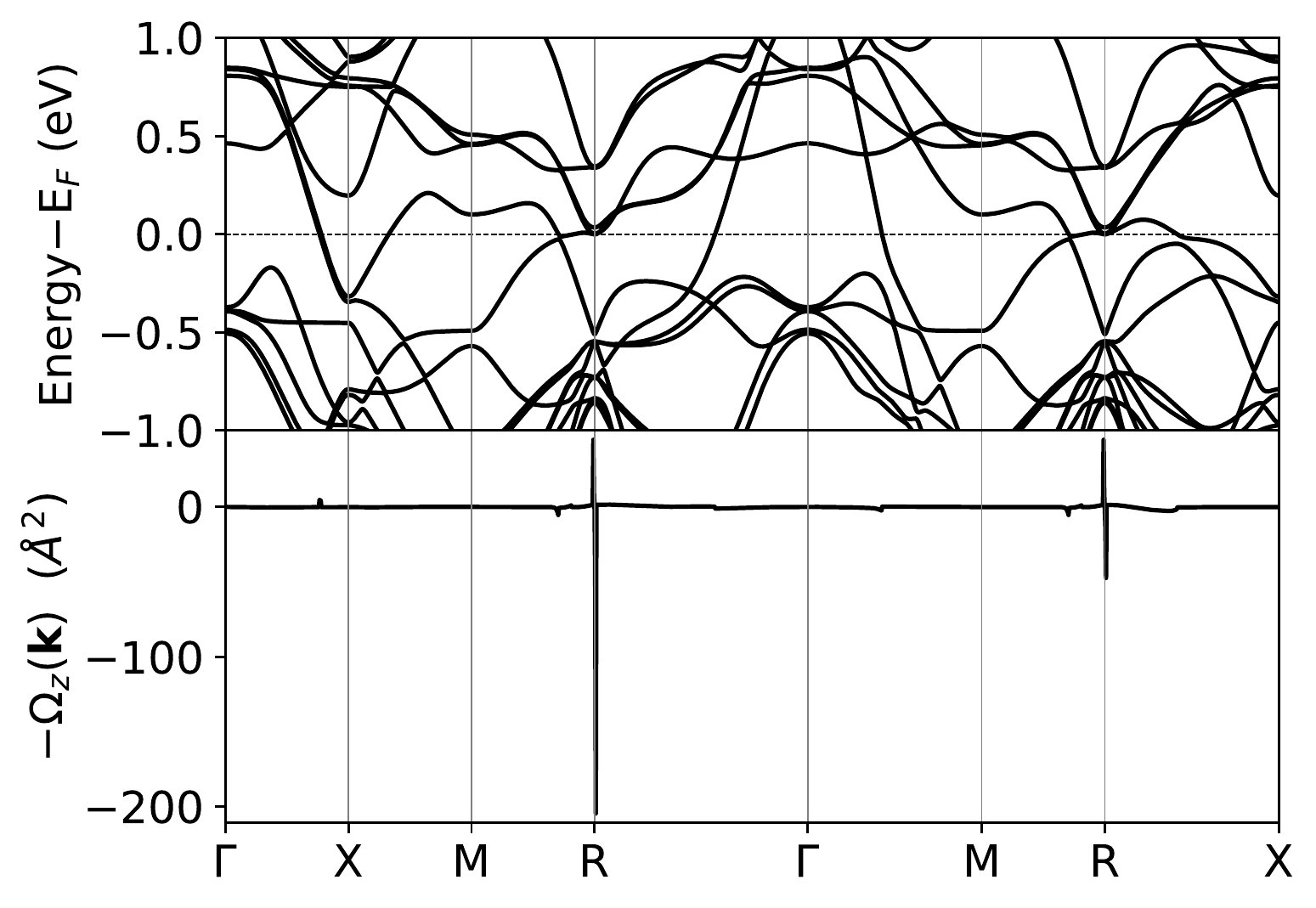}
 \caption{\textbf{DFT calculations of $\Gamma_{\rm 4g}$ Mn$_3$NiN.} Band structure (top) and Berry curvature summed over filled bands (bottom) along high-symmetry lines of the Brillouin zone.}
 \label{fig:4}
 \end{figure}

As already mentioned in the introduction, the $\Gamma_{4g}$ order of Mn$_3$NiN is essentially identical to the order of Mn moments in cubic Mn$_3$Ir and Mn$_3$Pt, and for the same symmetry reason as in the latter cases it should have the AHE. In contrast, the $\Gamma_{\rm 5g}$ state has a cubic symmetry that forbids AHE. To get a more quantitative estimate of its size in pure Mn$_3$NiN, which may allow us to make connections with the Cu-doped samples, we performed DFT calculations of the intrinsic AHC of the $\Gamma_{\rm 4g}$ phase of Mn$_3$NiN\cite{Mn311SM, Giannozzi_2009, wannier90, Hamann_2013}.
%using Quantum ESPRESSO \cite{Giannozzi_2009} and Wannier90 \cite{wannier90}.
%Fully-relativistic norm-conserving pseudopotentials are generated using the ONCVPSP code \cite{Hamann_2013}. The experimental lattice constant of 3.89 \AA is adopted. Cutoffs of 100Ry and 400Ry are used for wavefunctions and charge densities in the self-consistent calculations. A momentum space mesh of $8\times 8 \times 8$ is used for the DFT calculation and the wannierisation, and that of $150\times 150\times 150$ with $5\times 5\times 5$ adaptive refinement is used when calculating the anomalous Hall conductivity.
The band structure and Berry curvature summed over filled bands plotted along high-symmetry lines in the Brillouin zone are shown in Fig.~\ref{fig:4}. We find that the local spin moment on each Mn is 2.7 $\mu_B$ with a net moment of $\sim 0.007 \mu_B$/Mn, qualitatively agreeing with the experimental results. However, the calculated $\sigma_{\rm AH} = 525$ ($\Omega$ cm)$^{-1}$ (along the (111) direction) is much larger than the experimental value. It is well known that the intrinsic and extrinsic contributions to the AHE are difficult to be separated in experiments unless the the systems shows quantum AHE, in which only the intrinsic contribution survives. Therefore the intrinsic contribution alone is not usually enough for comparison with experiments. Nonetheless, the large difference between the calculated results and the experimental value is likely to be due to the modification of the electronic structure near Fermi surface by the significant amount of Cu doping. The neutron measurements indicate that right below 260K, Mn$_{3}$NiN system is in the $\Gamma_{\rm 4g}$ state\cite{Mn3NiN2013JAP}. But the Mn moments rotate toward the $\Gamma_{\rm 5g}$ state with decreasing temperature, which becomes dominant below 100K\cite{Mn3NiN2013JAP}. The Cu doping is therefore essential in stabilizing the $\Gamma_{\rm 4g}$ order at low temperatures which allows AHE. However, its quantitative influence on the AHE in the $\Gamma_{\rm 4g}$ state may depend on many system specific details and will be studied in the future.

In conclusion, we have shown that Cu-doped Mn$_{3}$NiN system is a noncollinear antiferromagnet having the AHE. Compared with intermetallic binary Mn compounds\cite{Mn3Pt2018NE}, it is much easier to obtain high quality epitaxial Mn$_{3}$Ni$_{1-x}$Cu$_{x}$N film on oxide substrates, giving much advantage for the preparation of multilayer devices in future. Combined with other attractive properties of the antiperovskite nitride systems, our finding may open a new avenue for their application in antiferromagnetic spintronics.

\acknowledgments
${Acknowledgments}$: The authors would like to thank German Hammerl, Sven Esser, Maximilian Uhl, and Anton Jesche for helpful discussion and experimental attempt. The work in Augsburg was supported by the German Science Foundation through the priority program SPP 1666. The work in Nagoya was supported by JSPS KAKENHI Grant Number 17K17801 and Kato foundation for Promotion of Science. T. H. also acknowledges Research Grant for Young Scientists (Research Center for Materials Backcasting Technology, School of Engineering, Nagoya University). This work utilized the RMACC Summit supercomputer, which is supported by the National Science Foundation (awards ACI-1532235 and ACI-1532236), the University of Colorado Boulder and Colorado State University. The RMACC Summit supercomputer is a joint effort of the University of Colorado Boulder and Colorado State University.

${Note}$ ${added}$: During the preparation and submission of our manuscript, we became aware of two papers about the AHE in Mn$_{3}$AN system\cite{Mn3AN2019AHE, Mn3NiN2019AHE}. The experimental paper reported the observation of AHE in Mn$_{3}$NiN film, stabilized by the strain between film and substrate\cite{Mn3NiN2019AHE}. Meanwhile, we adopt a totally different strategy, namely Cu substituting on Ni site in Mn$_{3}$Ni$_{1-x}$Cu$_{x}$N to stabilize the AHE.

\bibliography{Mn311}

\end{document}